\documentclass{SciPost}

\hypersetup{
  colorlinks,
  linkcolor={red!50!black},
  citecolor={blue!50!black},
  urlcolor={blue!80!black},
  pdftitle={Localized anomaly detection with a calorimeter normalizing flow},
  pdfauthor={Raunav Mendiratta, Jason Weitz, Javier Duarte}
}

\usepackage[T1]{fontenc}
\usepackage[bitstream-charter]{mathdesign} 
\usepackage{booktabs}

\DeclareSymbolFont{usualmathcal}{OMS}{cmsy}{m}{n}
\DeclareSymbolFontAlphabet{\mathcal}{usualmathcal}
\usepackage{siunitx}
\graphicspath{{./}}
\fancypagestyle{SPstyle}{
\fancyhf{}
\lhead{\colorbox{scipostblue}{\bf \color{white} ~SciPost Physics }}
\rhead{{\bf \color{scipostdeepblue} ~Submission }}

\fancyfoot[C]{\textbf{\thepage}}
}

\newcommand{\Erec}{E^{\mathrm{rec}}}
\newcommand{\Mmax}{M}
\newcommand{\zII}{z}
\newcommand{\sG}{s_G}
\newcommand{\ddec}{d_{\mathrm{dec}}}
\newcommand{\phat}{\hat p}

\begin{document}

\pagestyle{SPstyle}

\begin{center}{\Large \textbf{\color{scipostdeepblue}{
Localized anomaly detection with a calorimeter normalizing flow}}}\end{center}

\begin{center}
    \textbf{Raunav Mendiratta\textsuperscript{1, 2},
            Jason Weitz\textsuperscript{2$\star$} and
            Javier Duarte\textsuperscript{2}}
\end{center}

\begin{center}
    {\bf 1} Westview High School, San Diego, USA \\
    {\bf 2} Department of Physics, University of California San Diego, La Jolla, USA \\
    [\baselineskip]
    $\star$ \href{mailto:jdweitz@ucsd.edu}{\small jdweitz@ucsd.edu}
\end{center}

\section*{\color{scipostdeepblue}{Abstract}}
\textbf{\boldmath{
Normalizing flows trained as fast calorimeter simulators also give the likelihood of each shower, which can serve as an anomaly score.
Because this likelihood sums over every voxel, a signal confined to a few voxels can be lost in the fluctuations of the rest of the shower.
For example, in a public benchmark dataset in which a long-lived neutral particle decays to two photons, the CaloFlow likelihood used in previous work selects less than $5\%$ of the signal at a background efficiency of $10^{-3}$ when the particle has the largest simulated mass and decays at or just behind the front face of the calorimeter.
In this work, we define a regional anomaly score by summing the squared latent coordinates of the same flow within small detector regions around the point where the particle enters the calorimeter, with the threshold set using background showers alone.
At the same background efficiency, this regional anomaly score selects more than $95\%$ of the signal.
The per-coordinate likelihood terms, summed over the same regions, perform equally well, but the squared latent coordinates have a known distribution for background showers and clearly illustrate when a shower departs from it.
Therefore, normalizing flows may be sensitive to new physics, even if the global likelihood is not.
}}

\vspace{\baselineskip}

\noindent\textcolor{white!90!black}{%
\fbox{\parbox{0.975\linewidth}{%
\textcolor{white!40!black}{\begin{tabular}{lr}%
  \begin{minipage}{0.6\textwidth}%
    {\small Copyright attribution to authors. \newline
    This work is a submission to SciPost Physics. \newline
    License information to appear upon publication. \newline
    Publication information to appear upon publication.}
  \end{minipage} & \begin{minipage}{0.4\textwidth}
    {\small Received Date \newline Accepted Date \newline Published Date}%
  \end{minipage}
\end{tabular}}
}}
}

% \linenumbers

\vspace{10pt}
\noindent\rule{\textwidth}{1pt}\par
\tableofcontents
\noindent\rule{\textwidth}{1pt}\par
\vspace{10pt}

\section{Introduction} \label{sec:intro}

Searches at the CERN Large Hadron Collider (LHC) are usually optimized for specific signal models.
Unsupervised anomaly detection instead builds an event score from background data alone~\cite{Kasieczka:2021LHCOlympics,Karagiorgi:2021MLBSM,Belis:2024Review,Farina:2018Autoencoders,Heimel:2018QCDorWhat}, and resonant methods estimate the background from the sidebands of a known variable~\cite{Collins:2018CWoLa,Nachman:2020ANODE}.
Generative models that learn the background probability density give such a score as the likelihood of each event~\cite{Mikuni:2023Diffusion} or as another quantity computed from the trained model~\cite{Vaselli:2026FAD}.

Generative models are also being developed as fast surrogates for calorimeter shower simulation, which takes a large share of the computing needed for LHC detector simulation~\cite{Krause:2024CaloChallenge,Hashemi:2024GenReview}.
A \emph{normalizing flow} maps a shower, represented by the energies deposited in the calorimeter cells or voxels, through an invertible transformation to Gaussian latent coordinates.
Unlike most other generative models, it provides the exact likelihood of any shower and can generate new ones~\cite{Papamakarios:2019NormalizingFlows}, so a flow trained for simulation can be reused for inference~\cite{Du:2024UnifyingSimInference} and anomaly detection~\cite{Krause:2024AD}.
Reference~\cite{Krause:2024AD} used the likelihood of CaloFlow~\cite{Krause:2021CaloFlow,Krause:2021CaloFlowII} to separate the showers of a long-lived neutral particle $X$ that decays to two photons from single photons.
This signature is relevant to searches for displaced or collimated photons~\cite{ATLAS:2023DiphotonDisplaced,ATLAS:2024ALP4gamma} and for long-lived particles that decay in the calorimeters~\cite{ATLAS:2024LLPCalo}.
Based on the greatest improvement in significance $\varepsilon_S/\sqrt{\varepsilon_B}$ over all thresholds, the likelihood was the least sensitive in two cases.
For the three lightest masses, up to \SI{0.5}{GeV}, decays in front of the calorimeter or at its front face were difficult to detect, which Ref.~\cite{Krause:2024AD} attributed to highly collimated photon pairs whose showers resemble those of single photons~\cite{CMS:2023MergedDiphotons}.
For the heaviest mass, \SI{5}{GeV}, decays at the front face were less anomalous than decays farther upstream, because the two photons have little distance to separate before showering.
At a fixed background efficiency of $10^{-3}$, the likelihood selects less than $5\%$ of the \SI{5}{GeV} decays at or just behind the front face (Sec.~\ref{sec:results:near}).
Reference~\cite{Krause:2024AD} also trained supervised classifiers on the voxel energies.
Each did better than the likelihood on the mass it was trained on, but the classifier trained on the lightest mass lost much of its sensitivity to the heaviest.

For these \SI{5}{GeV} decays, and for upstream decays of \SI{0.5}{GeV} particles, the flow has the information needed to find the signal, but the likelihood combines it poorly.
The likelihood sums contributions from all voxels of the shower.
When the two photons overlap, the signal shower differs from a photon shower mainly in a few voxels near the shower core, and this difference is diluted by the fluctuations of the many other voxels.
The situation resembles the choice between a global goodness-of-fit test and a bump hunt in a narrow mass window.
Restricting the test to a small region makes it more sensitive to a signal there, at the cost of a trials factor for the number of regions searched, the look-elsewhere effect~\cite{Gross:2010LEE,Hein:2025LookEverywhere}.

We therefore build a \emph{regional} anomaly score from the latent representation of CaloFlow.
Its second flow, Flow-II, maps each shower to one latent coordinate per voxel, and its likelihood contains the sum of the squares of all of them.
Without retraining the flow, we assign each latent coordinate to its voxel and sum the squares within detector regions centered on the point where the particle enters the calorimeter, which we call the impact position.
The largest standardized sum is the score.
Its threshold is set on background photons alone with conformal calibration, which controls the fraction of background showers selected, the background efficiency.
For a particle of mass \SI{5}{GeV} that decays at or just behind the front face, the regional score selects almost all of the signal, while the likelihood selects only a few percent at the same background efficiency.
It also selects most of the upstream decays of \SI{0.5}{GeV} particles, which the likelihood misses.
The per-coordinate terms of the Flow-II likelihood, summed over the same regions, perform equally well, so the gain comes from localizing the score.

That a likelihood can be a poor anomaly score is known in machine learning. Deep generative models can assign higher likelihoods to unfamiliar inputs than to their own training data~\cite{Nalisnick:2019KnowWhat}, an effect that has been attributed to low-level input statistics~\cite{Kirichenko:2020FlowsOOD,Serra:2020InputComplexity,Ren:2019LikelihoodRatio,Schirrmeister:2020HierarchicalOOD}.
Even a perfect density model does not guarantee anomaly detection, because the density depends on how the input is represented~\cite{LeLan:2021PerfectDensity}. In collider physics, density-based and autoencoder-based scores likewise favor some anomalies over others~\cite{Kasieczka:2022CoordinateAD,Buss:2022Anomalous,Finke:2021Autoencoders,Fraser:2022Challenges}.
Proposed alternatives include typicality tests~\cite{Nalisnick:2020Typicality}, model ensembles~\cite{Choi:2018WAIC}, and goodness-of-fit tests on flow latent coordinates~\cite{Jiang:2022RevisitingFlows,Bomatter:2026SITN}.
In industrial inspection, defects are located with per-position maps built from patch features or from the latent coordinates and likelihoods of flows~\cite{Defard:2021PaDiM,Roth:2022PatchCore,Gudovskiy:2022CFLOWAD,Rudolph:2022CSFlow,Zhou:2025MSFlow}, and U-Flow~\cite{Tailanian:2024UFlow} scores an image by the maximum of a map of squared latent coordinates.
U-Flow also locates anomalous regions with a $\chi^2$ test on these coordinates, with a threshold on the expected number of false alarms among all regions tested.
We apply this idea to a calorimeter flow whose latent coordinates are not arranged on the detector grid.
We calibrate the score with split-conformal $p$-values for outlier detection~\cite{Vovk:2005ALRW,Bates:2023ConformalPvalues}, separately in bins of reconstructed energy~\cite{Vovk:2012ConditionalValidity}.
Conformal $p$-values have also been used in high-energy physics~\cite{Araz:2025ConformalHEP,Araz:2026LookElsewhere}.

Section~\ref{sec:model} describes the benchmark and the CaloFlow likelihood, Sec.~\ref{sec:score} defines the regional score and its calibration, Sec.~\ref{sec:results} presents the results, and Sec.~\ref{sec:conclusions} concludes.

\section{Dataset and model} \label{sec:model}

\subsection{Displaced-diphoton benchmark} \label{sec:model:data}

We use the public displaced-diphoton benchmark of Refs.~\cite{Pang:2023ADwithCF,Krause:2024AD}, simulated with Geant4~\cite{Agostinelli:2003Geant4} in a three-layer electromagnetic sampling calorimeter based on the CaloGAN detector~\cite{Paganini:2018CaloGAN_PRL,Paganini:2018CaloGAN_PRD}.
The three layers, numbered 0 to 2 from the front, have $288$, $144$, and $72$ voxels, $504$ in total, and Appendix~\ref{app:impl:data} gives the geometry.
Every particle is emitted from the origin along the calorimeter axis, which meets the front face at the corner shared by the four central voxels of layer~1.
This point is the impact position introduced in Sec.~\ref{sec:intro}.

The background consists of two independent samples of $10^5$ photons with energies distributed uniformly between $1$ and \SI{100}{GeV}.
We train CaloFlow five times, with different random seeds, on the first sample and use the second for design, calibration, and testing (Sec.~\ref{sec:score:calib}).
In the signal samples, a neutral particle $X$ of mass $m_X$ and energy $E_X=\SI{50}{GeV}$ decays to two photons at a distance $\ddec$ from the origin, for the $28$ combinations
\begin{equation}
 m_X\in\{5,0.5,0.05,0.005\}\,\si{GeV},\qquad
 \ddec\in\{0.33,0.66,1.00,1.04,1.08,1.16,1.24\}\,\si{m}.
 \label{eq:grid}
\end{equation}
Reference~\cite{Krause:2024AD} denotes the particle by $\chi$ and the decay position by $z$, a symbol we use for the latent coordinates of the flow.
The front face of the calorimeter is at \SI{1.00}{m}.
We call decays at $\ddec=\SI{0.33}{m}$ and \SI{0.66}{m} upstream decays and those at \SI{1.00}{m} and \SI{1.04}{m} front-face decays.
Sections~\ref{sec:results:near} and \ref{sec:results:mech} study the two front-face samples with $m_X=\SI{5}{GeV}$.

\begin{figure}[!tb]
\centering
\includegraphics[width=\linewidth]{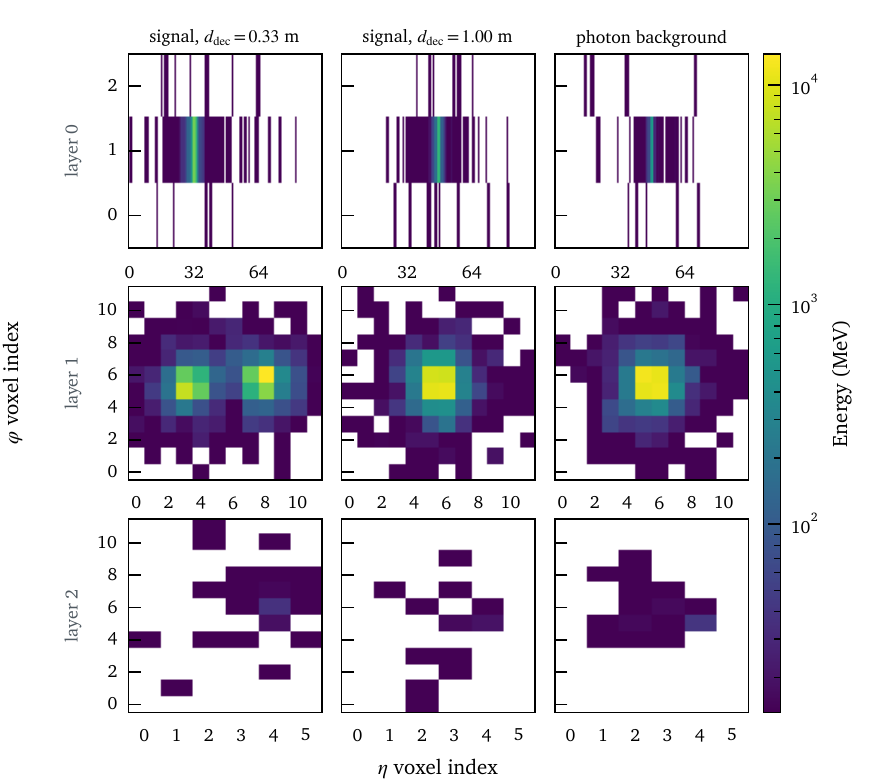}
\caption{Energy deposited in the three calorimeter layers (rows) by individual showers. The columns show a particle with $m_X=\SI{5}{GeV}$ decaying upstream at $\ddec=\SI{0.33}{m}$, a particle of the same mass decaying at the calorimeter front face at $\ddec=\SI{1.00}{m}$, and a background photon. The upstream decay produces two separate cores, while the front-face decay produces a single core that is hard to tell from that of the photon. The axes give the voxel indices in the two transverse directions, labeled $\varphi$ and $\eta$ as in Ref.~\cite{Krause:2024AD}. All panels use the same logarithmic energy scale; white voxels contain no deposited energy.}
\label{fig:showers}
\end{figure}

The opening angle between the two photons is at least
\begin{equation}
 \theta_{\gamma\gamma}=2\arcsin\!\left(\frac{m_X}{E_X}\right),
 \label{eq:opening}
\end{equation}
which for $m_X=\SI{5}{GeV}$ is \SI{0.2}{rad}. Each smaller mass in Eq.~\eqref{eq:grid} reduces the angle by about a factor of ten.
When a particle with $m_X=\SI{5}{GeV}$ decays upstream, the two photons enter the calorimeter far apart and produce two separate cores (Fig.~\ref{fig:showers}, left).
When it decays at the front face, their axes enter layer~1 only about \SI{2}{cm} apart, half the width of a layer-1 voxel, and the two showers overlap (center).

\subsection{CaloFlow likelihood and latent space} \label{sec:model:rep}

CaloFlow factorizes the shower density into a flow for the three layer energies $E_0$, $E_1$, and $E_2$, called Flow-I, and a flow for the shower image $\hat I$ normalized to the energy of each layer, called Flow-II.
Both are conditioned on the incident energy $E_{\mathrm{inc}}$, and the model density is
\begin{equation}
 p_{\mathrm{CF}}(E_0,E_1,E_2,\hat I\mid E_{\mathrm{inc}})
 =p_{\mathrm{I}}(E_0,E_1,E_2\mid E_{\mathrm{inc}})\,
 p_{\mathrm{II}}(\hat I\mid E_0,E_1,E_2,E_{\mathrm{inc}}).
 \label{eq:factorization}
\end{equation}
In an application $E_{\mathrm{inc}}$ is unknown and is replaced by its estimate $\Erec$ from the layer energies (Appendix~\ref{app:impl:data}).
We use $-\log p_{\mathrm{CF}}$, the anomaly score of Ref.~\cite{Krause:2024AD} with the Flow-I preprocessing of Appendix~\ref{app:impl:flow1}, as the likelihood score and baseline, and call $-\log p_{\mathrm{I}}$ the Flow-I term.

Flow-II maps the preprocessed image through an invertible transformation $f$ to latent coordinates $\zII\in\mathbb R^D$, with $D=504$ and a standard normal base density. Leaving out the conditioning, the change-of-variables formula of normalizing flows~\cite{Papamakarios:2019NormalizingFlows,Krause:2021CaloFlow} gives
\begin{equation}
 -\log p_{\mathrm{II}}=\frac{D}{2}\log(2\pi)
 +\frac{1}{2}\lVert\zII\rVert^2-\log\left|\det J_f\right|,
 \label{eq:decomp}
\end{equation}
where $J_f$ is the Jacobian of $f$.
The first two terms are the negative logarithm of the standard normal density at $\zII$, which grows with the distance of $\zII$ from the origin.
The last term accounts for how strongly $f$ stretches or compresses the space of images near the shower.
We call $-\log p_{\mathrm{II}}$ the image likelihood, the last term the log-Jacobian term, and $\lVert\zII\rVert^2$ the global latent norm. All scores in this paper are oriented so that a higher value is more anomalous.
Because Flow-II is autoregressive, the log-Jacobian term is exactly a sum of one contribution $-\ell_a$ per latent coordinate $\zII_a$ (Appendix~\ref{app:impl:regional-likelihood}), so that, up to a constant, $-\log p_{\mathrm{II}}$ is a sum of the per-coordinate likelihood terms $\tfrac12\zII_a^2-\ell_a$.

\section{Regional anomaly score} \label{sec:score}

\subsection{Detector regions} \label{sec:score:regions}

The regional score needs each Flow-II latent coordinate to be assigned to a voxel.
We obtain the assignment by following each coordinate through the fixed permutations of the flow, which act separately within each layer.
Because the flow is autoregressive, a latent coordinate also depends on other voxels, but for the four central layer-1 coordinates most of the dependence is on those four voxels (Appendix~\ref{app:impl:regions}).

\begin{figure}[!tb]
\centering
\includegraphics[width=0.94\linewidth]{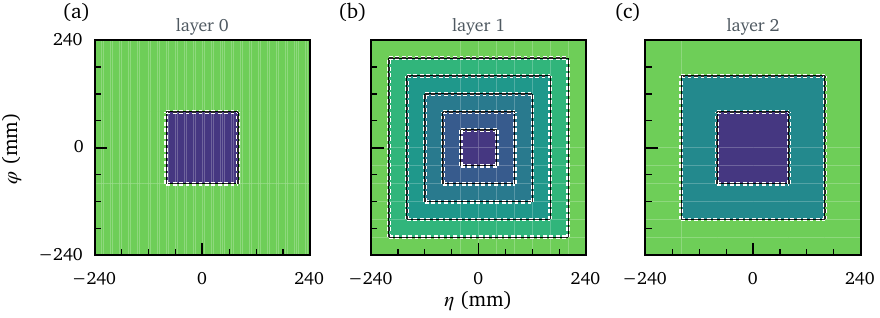}
\caption{Concentric square regions in (a) layer~0, (b) layer~1, and (c) layer~2. The impact position is at the origin of each panel. Together with the three complete layers, these regions form the $14$ regions of the regional score. Dashed lines show the region boundaries, and darker shading denotes regions closer to the impact position. All panels cover the same $\SI{480}{mm}\times\SI{480}{mm}$ transverse area.}
\label{fig:regions}
\end{figure}

Photons arrive along the calorimeter axis, and their showers are symmetric about it.
In each layer we therefore use a central square and square rings around it, centered on the impact position and bounded by voxel edges, together with the three complete layers (Fig.~\ref{fig:regions}).
This gives $14$ regions, chosen from the detector geometry and the symmetry of photon showers without using the signal.
The smallest region contains the four layer-1 voxels around the impact position.
Appendix~\ref{app:impl:regions} gives the construction.

For a region $G$ of $|G|$ voxels, we sum the squared latent coordinates assigned to its voxels and standardize the sum with the mean and standard deviation of a $\chi^2$ distribution with $|G|$ degrees of freedom,
\begin{equation}
 Q_G=\sum_{a\in G}\zII_a^2,\qquad
 \sG=\frac{Q_G-|G|}{\sqrt{2|G|}},\qquad
 \Mmax=\max_G\sG.
 \label{eq:M}
\end{equation}
The maximum runs over the $14$ regions, and we call $\Mmax$ the regional score.
Like a scan statistic~\cite{Kulldorff:1997ScanStatistic,AriasCastro:2018DistributionFree}, it is a maximum over spatial windows.

The per-coordinate likelihood terms, summed over regions in the same way, perform as well (Sec.~\ref{sec:results:near}).
We use the squared latent coordinates because they have a known distribution for photons.
The flow is trained to map photon showers to independent standard normal coordinates, so every voxel has the same reference, and a large $\zII_a^2$ marks a voxel that the model does not expect for a photon.
The values for all voxels form an anomaly map that shows where in the detector a shower departs from the background (Sec.~\ref{sec:results:near}), similar to the maps of flow-based defect detectors~\cite{Rudolph:2022CSFlow,Zhou:2025MSFlow,Tailanian:2024UFlow}.
The same reference sets the standardization in Eq.~\eqref{eq:M}.
Under the model each $Q_G$ follows a $\chi^2$ distribution, and each $\sG$ is a goodness-of-fit statistic for the latent coordinates of its region~\cite{Jiang:2022RevisitingFlows,Bomatter:2026SITN}.
For the test photons this holds in all $14$ regions except far in the tail (Appendix~\ref{app:impl:regions}), so we take the threshold from the photons themselves rather than from the $\chi^2$ distribution.
The log-Jacobian contributions have no such reference.
Their split among the coordinates follows the internal order of the flow, so the sum of the likelihood terms over a region, the regional likelihood of Sec.~\ref{sec:results:near}, is not the likelihood of the voxels in that region, although it can still serve as a score.

\subsection{Conformal calibration} \label{sec:score:calib}

We split the second photon sample once into a design sample, which fixes the five bins of reconstructed energy (Appendix~\ref{app:impl:calib}) and the standardizations of the scores other than $\Mmax$, a calibration sample, which sets the thresholds, and a test sample, on which the background efficiency is measured.
A shower $x$ in energy bin $b$ is ranked against the $n_b$ calibration photons $\mathcal C_b$ of the same bin.
Its conformal $p$-value is
\begin{equation}
 \phat(x)=\frac{1+\sum_{j\in\mathcal C_b}\mathbf 1[\Mmax(x_j)\ge \Mmax(x)]}{n_b+1},
 \label{eq:pvalue}
\end{equation}
where $\mathbf 1[\cdot]$ is the indicator function.
If the score and the bins are fixed independently of the calibration and test showers, and if calibration and future background showers are exchangeable within a bin, selecting the showers with $\phat\le\alpha$ has a false-positive probability of at most $\alpha$ in every bin, where the probability is taken also over the draw of the calibration sample.
This binned form is known as Mondrian conformal prediction~\cite{Vovk:2005ALRW,Vovk:2012ConditionalValidity}.

We compute $\phat$ from $\Mmax$ after the maximization over regions, so the calibration accounts for the number of regions and their correlations, as in the single-step max-$T$ procedure~\cite{Westfall:1993MaxT}. A related method, Max-Rank~\cite{Timans:2025MaxRank}, calibrates the maximum of the per-region ranks instead of the maximum of the scores.
The other scores are calibrated in the same way, and Appendix~\ref{app:impl:calib} checks the background efficiency achieved by the regional score.

Because the conformal $p$-values are discrete, the threshold $\phat\le10^{-3}$ selects slightly different background fractions for different scores.
To compare scores at nearly the same background efficiency, we instead threshold each one at the $10^{-3}$ quantile of its $p$-values on the test photons and report the signal efficiency $\varepsilon_S$, the fraction of signal showers selected, at a background efficiency $\varepsilon_B=10^{-3}$.
This matched threshold carries no finite-sample guarantee.
The conformal threshold is used in Sec.~\ref{sec:results:mech} and in Appendices~\ref{app:impl:regions} and \ref{app:impl:calib}.
Values and uncertainties are means and standard deviations over the five trained flows, evaluated on the same test photons.

\section{Results} \label{sec:results}

\subsection{Front-face samples} \label{sec:results:near}

Table~\ref{tab:eff} lists the signal efficiencies for the two front-face samples.
The likelihood score selects less than $5\%$ of the signal, and the image likelihood, without the Flow-I term, does no better.
The global latent norm, which drops the log-Jacobian term but still sums over all coordinates, selects a little under half.
The regional score selects more than $95\%$.

\begin{table}[!htb]
\centering
\caption{Signal efficiency at $\varepsilon_B=10^{-3}$ for the two front-face samples ($m_X=\SI{5}{GeV}$, $\ddec=\SI{1.00}{m}$ and \SI{1.04}{m}).
For each latent coordinate $a$, the term $\tfrac12 z_a^2$, its log-Jacobian contribution $-\ell_a$ (Appendix~\ref{app:impl:regional-likelihood}), or the per-coordinate likelihood term $\tfrac12 z_a^2-\ell_a$ is summed over all $504$ coordinates, over the three complete layers, or over the $14$ regions. In the last two cases the largest standardized sum defines the score. Only the sums over regions select nearly all of the signal.}
\label{tab:eff}
\small
\begin{tabular}{llrr}
\toprule
term per coordinate & summed over & $\varepsilon_S$, \SI{1.00}{m} & $\varepsilon_S$, \SI{1.04}{m} \\
\midrule
$\tfrac12 z_a^2$ & all coordinates (global latent norm) & $0.42 \pm 0.10$ & $0.46 \pm 0.15$ \\
 & three layers & $0.50 \pm 0.09$ & $0.65 \pm 0.13$ \\
 & regions (regional score $\Mmax$) & $0.964 \pm 0.005$ & $0.972 \pm 0.002$ \\
\midrule
$-\ell_a$ & all coordinates & $0.002 \pm 0.004$ & $0.006 \pm 0.007$ \\
 & regions & $0.86 \pm 0.07$ & $0.87 \pm 0.08$ \\
\midrule
$\tfrac12 z_a^2-\ell_a$ & all coordinates (image likelihood) & $0.006 \pm 0.009$ & $0.05 \pm 0.02$ \\
 & three layers & $0.03 \pm 0.04$ & $0.01 \pm 0.02$ \\
 & regions (regional likelihood) & $0.969 \pm 0.003$ & $0.979 \pm 0.002$ \\
\midrule
\multicolumn{2}{l}{likelihood score $-\log p_{\mathrm{CF}}$} & $0.006 \pm 0.006$ & $0.045 \pm 0.017$ \\
\bottomrule
\end{tabular}
\end{table}

Because the log-Jacobian term is a sum of the per-coordinate contributions $-\ell_a$ (Sec.~\ref{sec:model:rep}), it can be restricted to regions in the same way as the squared latent coordinates.
Summed over all coordinates, neither the log-Jacobian contributions nor the per-coordinate likelihood terms select more than a few percent of the signal.
Maximized over the regions, the log-Jacobian contributions alone select most of it, and the per-coordinate likelihood terms, a score we call the regional likelihood, perform as well as the regional score.
Sums over complete layers remain far less efficient.
The gain therefore comes from restricting the sum to a small region, not from the choice of terms.

\begin{figure}[!tb]
\centering
\includegraphics[width=\linewidth]{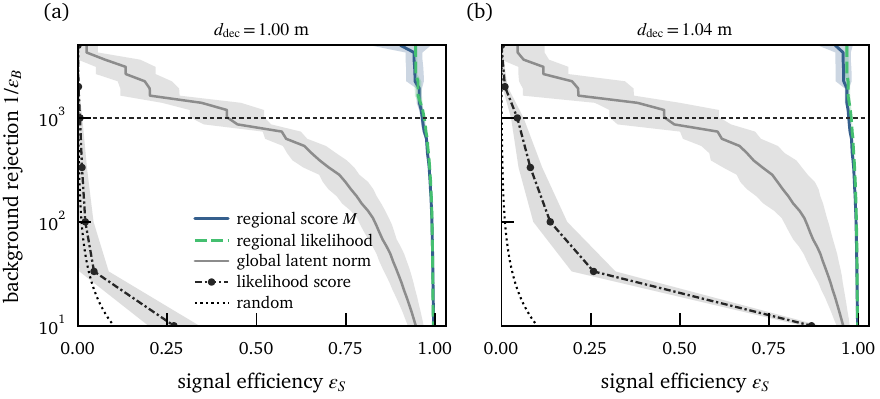}
\caption{Background rejection $1/\varepsilon_B$ as a function of the signal efficiency for the front-face samples at (a) $\ddec=\SI{1.00}{m}$ and (b) \SI{1.04}{m}. Each score is thresholded at the $\varepsilon_B$ quantile of its conformal $p$-values on the test photons. The dashed line marks $\varepsilon_B=10^{-3}$, the working point of Table~\ref{tab:eff}, and the dotted line a random selection, $\varepsilon_S=\varepsilon_B$. The regional score and the regional likelihood run close to $\varepsilon_S=1$ along the right edge of each panel. The likelihood score was evaluated only at the seven marked background efficiencies, the lowest of which lies on the upper edge of each panel. Bands show the standard deviation over the five trained flows.}
\label{fig:roc}
\end{figure}

Figure~\ref{fig:roc} extends the comparison to other background efficiencies.
The regional score and the regional likelihood nearly coincide, except at the lowest background efficiencies, where the regional likelihood is slightly higher.
Both stay above the global latent norm and the likelihood score over the whole range.

\begin{figure}[!tb]
\centering
\includegraphics[width=\linewidth]{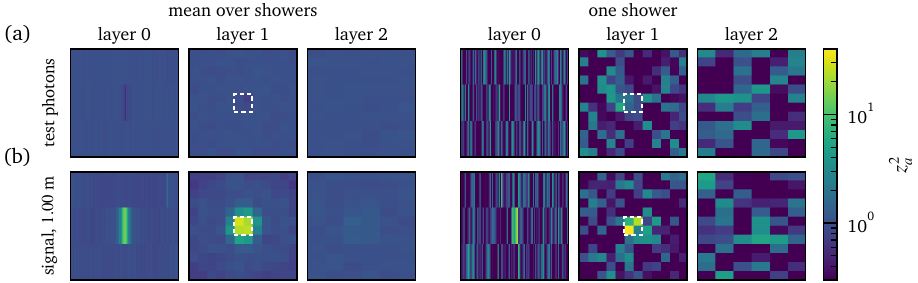}
\caption{Squared latent coordinate $\zII_a^2$ of each voxel for (a) test photons in the energy bin of the front-face samples and (b) signal showers with $m_X=\SI{5}{GeV}$ decaying at $\ddec=\SI{1.00}{m}$. The left block shows the mean over showers and over three of the five trained flows, and the right block shows, for one flow, the shower with the median regional score in each sample. Columns show the three layers, drawn to scale over the same area as in Fig.~\ref{fig:regions}, and the dashed square marks the innermost layer-1 region. For photons the mean is close to one almost everywhere and a single shower fluctuates about it, while for the signal the excess is concentrated at the impact position, on average and in a single shower.}
\label{fig:map}
\end{figure}

Figure~\ref{fig:map} shows the anomaly map, averaged over showers and for single showers.
For photons, the mean of $\zII_a^2$ is close to one in almost every voxel, as the model expects, and somewhat lower at the impact position.
For the signal it is more than ten times larger in the four voxels of the innermost layer-1 region and along the central strips of layer~0, at the impact position.
About half of the total excess over photons lies in those four layer-1 voxels.
A typical single signal shower, the one with the median regional score, shows the same excess at the impact position, while a typical single photon shows only scattered fluctuations.

The regional score depends on the assignment of latent coordinates to voxels.
Randomly permuting the assignment, which keeps the size of every region, lowers the efficiency of the regional score at every grid point with $m_X\ge\SI{0.5}{GeV}$ (Appendix~\ref{app:impl:regions}).
A sum of only the few largest squared latent coordinates, which ignores their position, recovers most but not all of the front-face signal.

Photon-identification variables, including the layer-0 strip variables that reject $\pi^0\to\gamma\gamma$ decays, select less than half of the signal when they are combined without labels (Appendix~\ref{app:impl:supervised}).
Trained with signal labels, a classifier on the same variables selects about as much as the regional score at $\varepsilon_B=10^{-3}$, and one on the latent coordinates selects slightly more.
The regional score, which uses no signal showers, thus comes close to these supervised references.

\subsection{Why the summed scores miss the signal} \label{sec:results:mech}

\begin{figure}[!tb]
\centering
\includegraphics[width=\linewidth]{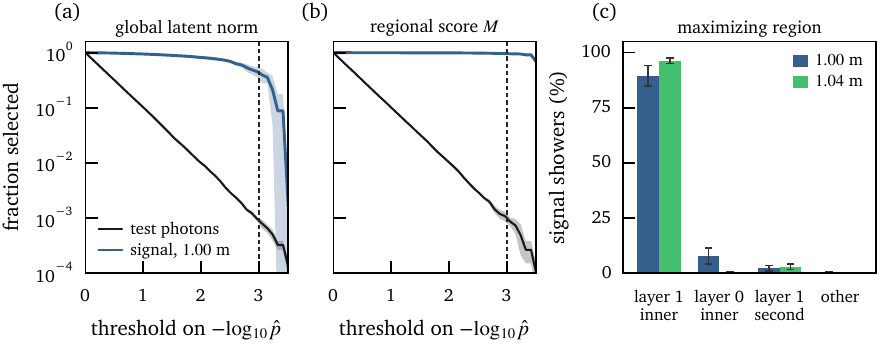}
\caption{Fraction of test photons and of signal showers with $m_X=\SI{5}{GeV}$ and $\ddec=\SI{1.00}{m}$ selected as a function of the threshold on $-\log_{10}\phat$, for (a) the global latent norm and (b) the regional score. The dashed line marks $\phat=10^{-3}$. For calibrated $p$-values the fraction of photons selected is close to $\phat$, so the photon curve follows the diagonal, and a score separates the signal when its signal curve stays high where the photon curve is low. Panel (c) gives the fraction of signal showers at \SI{1.00}{m} and \SI{1.04}{m} for which each region attains the maximum; ``inner'' and ``second'' denote the innermost region of a layer and the one surrounding it, and ``other'' collects the remaining regions. Bands and error bars show the standard deviation over the five trained flows.}
\label{fig:mech}
\end{figure}

The dilution can be seen in the separation between signal showers at \SI{1.00}{m} and photons, measured for each score as the difference of the medians in units of the standard deviation of the photons in the energy bin of the signal.
This separation is $33$ for the sum over the innermost layer-1 region, $5$ for the global latent norm, and $1$ for the image likelihood.
At $\varepsilon_B=10^{-3}$ the thresholds lie a few standard deviations above the photon median, so the sum over the innermost layer-1 region selects almost all of the signal, the global latent norm about half, and the image likelihood almost none.
Figure~\ref{fig:mech} shows, for the global latent norm and the regional score, the fractions of photons and signal showers selected as a function of the conformal threshold.
At $\phat\le10^{-3}$ both select about $10^{-3}$ of the photons.
For most signal showers the maximum of the regional score is attained in the innermost layer-1 region (Fig.~\ref{fig:mech}(c)).

The global latent norm compares the excess in four coordinates with the fluctuations of all $504$, the known weakness of sum-type tests against sparse signals~\cite{Donoho:2004HigherCriticism}.
The image likelihood adds the log-Jacobian term, whose spread among photons is about ten times that of $\tfrac12\lVert\zII\rVert^2$ (Appendix~\ref{app:impl:regions}).
Almost all of this spread comes from the log-Jacobian contributions of the layer-0 coordinates, which the signal at \SI{1.00}{m} changes only slightly.

\subsection{Dependence on mass and decay position} \label{sec:results:grid}

\begin{figure}[!tb]
\centering
\includegraphics[width=\linewidth]{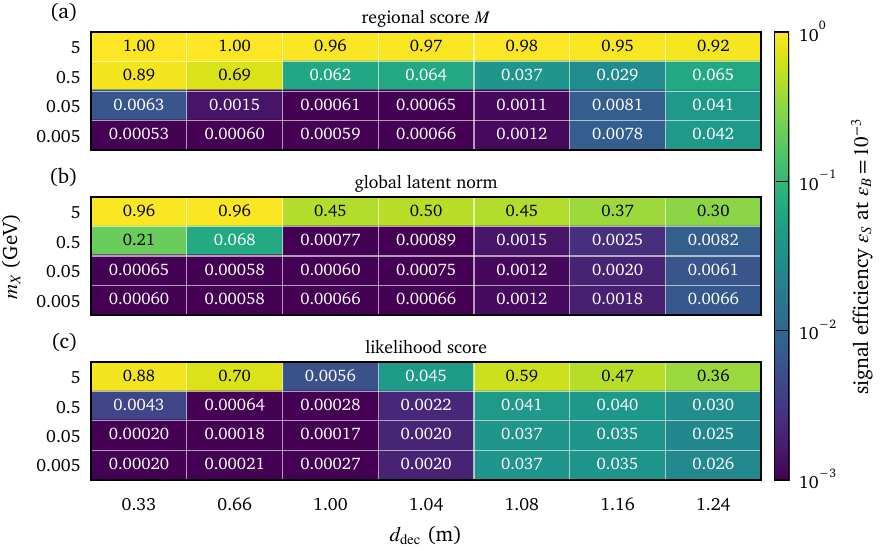}
\caption{Signal efficiency at $\varepsilon_B=10^{-3}$ as a function of the mass $m_X$ and the decay position $\ddec$ for (a) the regional score, (b) the global latent norm, and (c) the likelihood score $-\log p_{\mathrm{CF}}$, the baseline of Ref.~\cite{Krause:2024AD}. 
Each entry is the mean over the five trained flows. 
Panels (a) and (b) come from a separate evaluation of the flows compared to the one in Table~\ref{tab:eff}, and at the front-face points, the two agree within the spread over the flows.
The color scale is logarithmic and saturates at $10^{-3}$, the background efficiency, so the darkest entries are not separated from photons. 
The calorimeter front face is at $\ddec=\SI{1.00}{m}$.}
\label{fig:grid}
\end{figure}

Figure~\ref{fig:grid} shows the signal efficiency for all $28$ combinations of mass and decay position.
The efficiency of the regional score is set mainly by the separation of the two photon axes at the back of layer~1, which decreases with decreasing mass and with increasing $\ddec$.
It is low when the axes are less than about a centimeter apart, which is the case at every point with $m_X\le\SI{0.05}{GeV}$ and for the decays of \SI{0.5}{GeV} particles at or behind the front face.
It is high at every decay position for $m_X=\SI{5}{GeV}$ and for the upstream decays of \SI{0.5}{GeV} particles, whose axes are only about half a layer-1 voxel apart at the back of layer~1, so the score does not need two resolved showers.
For $m_X\le\SI{0.05}{GeV}$, the photon axes are at most a few millimeters apart, finer than the layer-0 strips, and no score selects upstream or front-face decays more than about six times as often as photons.

Reference~\cite{Krause:2024AD} reports the largest significance improvement $\varepsilon_S/\sqrt{\varepsilon_B}$ over all thresholds rather than $\varepsilon_S$ at a fixed $\varepsilon_B$, and the likelihood score in Fig.~\ref{fig:grid}(c) shows qualitatively the same dependence on mass and decay position as Fig.~3 of that reference.
Its efficiency does not follow the separation, and it misses the upstream decays of \SI{0.5}{GeV} particles, which the regional score selects.
For $m_X=\SI{5}{GeV}$ particles that decay at least \SI{8}{cm} inside the calorimeter, it selects a third or more of the signal, more than the global latent norm.
These decays occur near the back of layer~0 or behind it, so layer~0 receives little energy, and the signal moves the Flow-I term by several standard deviations of the photons, more for deeper decays (Appendix~\ref{app:impl:regions}). The regional score, built from Flow-II alone, does not use this information.
Across the grid, the regional likelihood selects nearly the same fraction of the signal as the regional score, slightly more for the two deepest \SI{5}{GeV} decays and slightly less for \SI{0.5}{GeV} particles decaying at \SI{0.66}{m} (Appendix~\ref{app:impl:regions}).

\section{Conclusions} \label{sec:conclusions}

The global likelihood of a generative model can miss a signal that changes only a small part of the detector.
For a neutral particle of mass \SI{5}{GeV} that decays to two photons at or just behind the calorimeter front face, the CaloFlow likelihood selects less than $5\%$ of the signal at a background efficiency of $10^{-3}$.
A regional score built from the latent coordinates of the same flow, without retraining, selects almost all of it at the same background efficiency.

The signal produces a large response in a few latent coordinates at the impact position.
The likelihood sums this response with the contributions of all other coordinates.
The sum dilutes it, so that even the global latent norm selects only about half of the signal, and the fluctuations of the layer-0 log-Jacobian contributions hide it almost completely.
Summed over the same regions, the per-coordinate terms of the image likelihood also recover the signal.

For anomaly detection with generative models, a likelihood that misses a signal is thus not evidence that the model is insensitive to it, and localized versions of the likelihood are worth including when anomaly scores are compared.
Without signal labels, the regional score selects about as much of the front-face signal as classifiers trained on these samples (Sec.~\ref{sec:results:near}).
A flow trained as a fast calorimeter simulator can be used in this way as long as its latent coordinates can be assigned to voxels, and other flow-based calorimeter simulators~\cite{Buckley:2024iCaloFlow,Ernst:2025CaloINN,Diefenbacher:2023L2LFlows} can be tested in the same way.
The anomaly map shows where a signal shower departs from the background, on average and for a single shower (Fig.~\ref{fig:map}), and the region that attains the maximum of the regional score indicates this location for each shower (Fig.~\ref{fig:mech}(c)).
Because the threshold is set on background showers, the background efficiency stays at or below the chosen level even though the trained flow departs from its Gaussian reference far in the tail.
In data, the threshold would be set in the same way on background showers from a control sample.
If that sample represents the background of the search, mismodeling by a flow trained on simulation affects the signal efficiency but not the background efficiency.

Every shower in the benchmark enters the calorimeter at the same point, an inherent limitation of the dataset, so the flows are trained and the regions defined at a single impact position.
Photon showers moved by two layer-1 voxels are almost all selected as anomalous, and smaller displacements cannot be emulated reliably by moving the voxel images, because the interpolation this requires changes the response of the flow by itself (Appendix~\ref{app:impl:regions}).
In a real detector the impact position varies, so the flow would have to be conditioned on it and the regions centered on its value reconstructed for each shower, which requires showers simulated at different impact positions.
The benchmark also uses a single signal energy and a simplified three-layer detector without electronic noise, and contains only single photons as background.
Since the regional score responds to small local differences, it also needs a flow that describes the core of the background showers well.
The regional score ignores the layer energies modeled by Flow-I, which change when the particle decays inside the calorimeter.
For particles with $m_X\le\SI{0.05}{GeV}$ decaying at \SI{1.08}{m} and \SI{1.16}{m}, it selects almost none of the signal and the likelihood score a few percent, so a combination with the Flow-I term could be tested there.
Particles light enough that the two photons are closer than the smallest voxels, here those with $m_X\le\SI{0.05}{GeV}$ decaying upstream or at the front face, are barely distinguished from photons by any of the scores.

\section*{Acknowledgements}
\emergencystretch=2em
This work used resources of the National Research Platform (NRP) at the University of California San Diego, which is supported in part by National Science Foundation (NSF) awards \mbox{CNS-1730158}, \mbox{ACI-1540112}, \mbox{ACI-1541349}, \mbox{OAC-1826967}, \mbox{OAC-2112167}, \mbox{CNS-2100237}, and \mbox{CNS-2120019}.

\paragraph{Author contributions}
Following the Contributor Roles Taxonomy (CRediT), R.~M. contributed to Conceptualization, Methodology, Software, Validation, Formal analysis, Investigation, Visualization, Writing (Original Draft), and Writing (Review \& Editing); J.~W. contributed to Supervision and Writing (Review \& Editing); and J.~D. contributed to Writing (Review \& Editing), Supervision, and Funding acquisition.
The authors acknowledge the use of generative AI writing assistance and coding assistance, and take full responsibility for the content of the publication.

\paragraph{Funding information}
This work was supported by the U.S. Department of Energy (DOE), Office of Science, Office of High Energy Physics under Award Number DE-SC0009919.
J.~W. was also supported by a Western Advanced Training for Computational High Energy Physics (WATCHEP) fellowship sponsored by the DOE, Office of Science, Office of High Energy Physics under Award Number DE-SC0023527.

\paragraph{Code and data availability}
The photon and signal samples are publicly available on Zenodo~\cite{Pang:2023ADwithCF}, and the analysis code is available at \url{https://github.com/raunavm/caloflow_ad} and archived on Zenodo~\cite{Mendiratta:2026Code}.

\begin{appendix}
\numberwithin{equation}{section}

\section{Implementation details} \label{app:impl}

\subsection{Detector, samples, and preprocessing} \label{app:impl:data}

The front face of the calorimeter, a $\SI{480}{mm}\times\SI{480}{mm}$ square, is located \SI{1.00}{m} from the origin, and the layer boundaries behind it lie at \SI{1.09}{m} and \SI{1.437}{m}.
Layers 0, 1, and 2 are divided into $3\times96$, $12\times12$, and $12\times6$ voxels.
In the two Cartesian transverse directions, labeled $\varphi$ and $\eta$ in Ref.~\cite{Krause:2024AD}, the voxel sizes are $(160,5)$~mm, $(40,40)$~mm, and $(40,80)$~mm.
The benchmark also contains decays at \SI{1.32}{m}, \SI{1.40}{m}, and \SI{1.44}{m}, deeper in the calorimeter, which we exclude because the median reconstructed energy of their $m_X=\SI{5}{GeV}$ samples is at most $59\%$ of $E_X$.
At the minimum opening angle, the two photon axes of a \SI{5}{GeV} particle decaying at \SI{1.00}{m} are \SI{18}{mm} apart at the entrance of layer~1 and \SI{88}{mm} apart at its exit, and for the two upstream positions they are \SI{153}{mm} and \SI{86}{mm} apart at the entrance of layer~1.

Since the incident energy is unknown in an application, the public code of Ref.~\cite{Krause:2024AD} reconstructs it from the layer energies as
\begin{equation}
 \Erec=\frac{E_0+E_1+E_2}{0.937},
 \label{eq:erec}
\end{equation}
where the constant is that of the code.
Reference~\cite{Krause:2024AD} describes the factor $1/0.937$ as the mean ratio of incident to deposited energy in the training sample, and in the second photon sample the mean ratio of the summed voxel energy to the incident energy is also $0.937$.
The flows are conditioned on the true incident energy during training and on $\Erec$ during evaluation.
Dividing each voxel energy by the total energy of its layer gives the normalized image $\hat I$.
As in Ref.~\cite{Krause:2024AD}, Flow-II models the logit of the normalized voxel energies $u$ after the mapping $u\mapsto\alpha+(1-2\alpha)u$ with $\alpha=10^{-6}$, and we use the architecture and training settings of the public code.
The first photon sample is split into $70\%$ for training and $30\%$ for validation, and each signal sample contains $10^5$ showers.
As in that code, noise drawn uniformly between $0$ and \SI{1}{keV} is added to every voxel in training and evaluation, similar to the dequantization of discrete data~\cite{Theis:2016Dequant}.
The likelihood score is evaluated in the preprocessed coordinates on which the flows are trained, and Flow-I uses the logarithmic preprocessing of Ref.~\cite{Krause:2024AD} (Appendix~\ref{app:impl:flow1}).

\subsection{Latent assignment, regions, and further checks} \label{app:impl:regions}

The fixed permutations of the public architecture act within blocks of $288$, $144$, and $72$ coordinates, one block per layer, so each latent coordinate can be followed to one voxel.
For the four latent coordinates assigned to the central layer-1 voxels, $86\%$ of the summed squared derivatives with respect to the voxel inputs of Flow-II, averaged over $512$ photons of the second sample, come from those four voxels.
For the next two square rings of layer~1 the fractions are $34\%$ and $50\%$.

For a voxel whose center is offset from the impact position by $(k_\varphi \Delta_\varphi,k_\eta \Delta_\eta)$, where $(\Delta_\varphi,\Delta_\eta)$ are the voxel sizes of its layer, the outer edge of the voxel lies at the radial coordinate
\begin{equation}
 r=\max\!\left[(|k_\varphi|+\tfrac12)\Delta_\varphi,
                (|k_\eta|+\tfrac12)\Delta_\eta\right].
 \label{eq:radius}
\end{equation}
The offsets $k$ are half-integers, except $k_\varphi$ in layer~0, where the impact position lies at a voxel center.
We place a region boundary at every value of $r$ that coincides with a voxel edge in both directions, at \SI{80}{mm} and \SI{240}{mm} in layer~0, at multiples of \SI{40}{mm} in layer~1, and at multiples of \SI{80}{mm} in layer~2.
This divides layers 0, 1, and 2 into two, six, and three regions, and the three complete layers bring the total to $14$.

For the test photons, averaged over the five flows, the mean of $\sG$ lies between $-0.16$ and $0.11$ and its standard deviation between $0.97$ and $1.09$ in every region, close to the values of $0$ and $1$ expected under the model.
Far in the tail the trained flow departs from this reference. The fraction of photons above the $\chi^2$ quantile at tail probability $10^{-3}$ is $1.5$ to $5.6$ times the expected value.

For each of five random permutations of the assignment of all $504$ latent coordinates to voxels, the efficiency of the regional score decreases at all $14$ points of Eq.~\eqref{eq:grid} with $m_X\ge\SI{0.5}{GeV}$, by a median factor of $2.5$ and by factors of $1.8$ and $1.6$ for the two front-face samples.
Without any assignment, the sum of the one, four, or eight largest squared latent coordinates selects at most $86\%$ and $87\%$ of the front-face samples.

Among the test photons in the energy bin of the front-face samples, the log-Jacobian term has a standard deviation of $153$ nats, compared with $15.8$ nats for $\tfrac12\lVert\zII\rVert^2$. The log-Jacobian contributions of the $288$ layer-0 coordinates together have a standard deviation of $150$ nats, compared with $0.4$ nats for the innermost layer-1 region, and the signal at \SI{1.00}{m} shifts their sum by $16$ nats on average over the five flows.
For $m_X=\SI{5}{GeV}$, the signal shifts the Flow-I term, averaged over the five flows, by $2.2$, $5.9$, and $11.6$ of its standard deviations among photons at $\ddec=\SI{1.08}{m}$, \SI{1.16}{m}, and \SI{1.24}{m}.

To compare the regional likelihood with the regional score over the $28$ points of Eq.~\eqref{eq:grid}, we evaluate the five flows a second time, with newly drawn dequantization noise, and compute both scores from the same evaluation.
The regional score reproduces Fig.~\ref{fig:grid} to within $0.02$ at every point.
In this evaluation the regional likelihood and the regional score differ in signal efficiency by at most $0.02$, except for $m_X=\SI{5}{GeV}$ at \SI{1.16}{m} and \SI{1.24}{m}, where the regional likelihood selects $0.97$ and $0.96$ against $0.94$ and $0.91$, and for $m_X=\SI{0.5}{GeV}$ at \SI{0.66}{m}, where it selects $0.63$ against $0.67$.

To test a displaced impact position, we move $20\,000$ of the test photon showers by \SI{80}{mm} along $\eta$.
This is a whole number of voxels in every layer, $16$, $2$, and $1$ in layers 0, 1, and 2, so the showers are moved exactly, apart from the energy that leaves the calorimeter at its edge.
With the threshold $\phat\le10^{-3}$ set on unshifted photons, the regional score selects more than $99.9\%$ of the moved photons for each of the five flows, so the trained flows and regions apply only at the impact position of the benchmark.
Smaller displacements are not a whole number of voxels in every layer, and moving a shower then requires interpolating between neighboring voxels, which smooths the image.
The interpolation alone changes the response of the flow.
Moving the showers by \SI{2.5}{mm} and back, with no net displacement, already raises the fraction selected to between $46\%$ and $61\%$, depending on the flow, so smaller displacements cannot be studied in this way.

\subsection{Calibration} \label{app:impl:calib}

The design, calibration, and test samples contain $20\,000$, $30\,000$, and $30\,000$ photons, and the remaining $20\,000$ photons of the second sample are not used by the conformal scores.
Together with the design sample, they train the photon-identification classifier of Appendix~\ref{app:impl:supervised}.
The five quantile bins of $\Erec$ defined on the design sample have inner edges at $20.5$, $40.3$, $60.2$, and \SI{80.0}{GeV}, and each bin holds about $6000$ calibration photons.
The front-face samples fall in the third bin.
Because the calibration sample is finite, the conformal $p$-values are discrete with steps of $1/(n_b+1)$, and the largest attainable level that does not exceed $10^{-3}$ ranges from $0.83\times10^{-3}$ to $1.00\times10^{-3}$ across the bins.
Over $40$ random partitions into calibration and test samples, the threshold $\phat\le10^{-3}$ on the regional score selects a fraction $(9.0\pm2.5)\times10^{-4}$ of the test photons, the mean and standard deviation over the partitions and the five flows.
A parametric $\chi^2$ threshold with a Bonferroni correction for the $14$ regions selects six times the nominal fraction.

\subsection{Flow-I preprocessing} \label{app:impl:flow1}

With the default settings of the public implementation of Ref.~\cite{Krause:2024AD}, Flow-I is trained on the logit transformation of the layer energies scaled to the unit interval, as in the original CaloFlow~\cite{Krause:2021CaloFlow}. The likelihood evaluation of that code instead always applies the logarithmic transformation $E_i\mapsto2[\log_{10}(E_i/\mathrm{MeV}+10^{-8})-1]$, which Ref.~\cite{Krause:2024AD} describes for Flow-I with an offset of \SI{1}{keV}.\footnote{In \texttt{run.py} of \url{https://github.com/Ian-Pang/AD_with_CF} at commit \texttt{6aec256}, Flow-I training applies the logarithmic transformation only with the non-default flag \texttt{-{}-log\_preprocess}, and the likelihood evaluation always applies it.} With the default settings, Flow-I is therefore evaluated in a different representation from the one on which it was trained. We trained Flow-I with the logarithmic transformation and left the architecture, optimizer, and training schedule unchanged.

\subsection{Decomposition of the image likelihood}
\label{app:impl:regional-likelihood}

Flow-II consists of eight masked autoregressive blocks~\cite{Papamakarios:2017MAF,Germain:2015MADE}. Each block transforms every coordinate with a monotonic rational-quadratic spline~\cite{Durkan:2019NSF} whose parameters depend on the preceding coordinates, and fixed permutations connect consecutive blocks. The Jacobian of a block is triangular, so its log-determinant is a sum of one term per coordinate. Tracking a coordinate through the permutations assigns its eight terms to the latent coordinate $a$ in which it ends. With $\ell_a$ denoting their sum, the image likelihood is
\begin{equation}
 -\log p_{\mathrm{II}}=\frac{D}{2}\log(2\pi)
 +\sum_a\left(\frac{1}{2}\zII_a^2-\ell_a\right).
 \label{eq:pervoxel}
\end{equation}
The extracted $\ell_a$ reproduce the log-determinant returned by the unmodified flow to within $5\times10^{-4}$ nats, and the tracked coordinate order reproduces the assignment of Sec.~\ref{sec:score:regions}. For each region $G$, the regional likelihood of Table~\ref{tab:eff} is built from $\sum_{a\in G}(\tfrac12\zII_a^2-\ell_a)$. We standardize each regional sum with the mean and standard deviation of the design photons in the same energy bin, take the maximum over the regions, and apply the calibration of Sec.~\ref{sec:score:calib}.

\subsection{Supervised reference classifiers} \label{app:impl:supervised}

The $13$ photon-identification variables are computed from the voxel energies.
They are the energy fractions in layers~0 and 2; the strip variables $E_{\mathrm{ratio}}$, $\Delta E_s$, $w_{s3}$, $w_{s,\mathrm{tot}}$, and $f_{\mathrm{side}}$ of ATLAS~\cite{ATLAS:2019PhotonID}, computed from the layer-0 $\eta$ profile summed over the three $\varphi$ rows; the ATLAS ratios $R_\eta$ and $R_\varphi$ and width $w_{\eta2}$ in layer~1, with windows of $2$ and $4$ voxels in place of $3$ and $7$ cells, centered on the impact position; and $\sigma_{i\eta i\eta}$ and $R_9$ of CMS~\cite{CMS:2015PhotonID}, together with the analogous width $\sigma_{i\varphi i\varphi}$.
Without labels, we score each variable alone with a two-sided conformal $p$-value and combine all $13$, standardized in each energy bin on the design photons, in a Mahalanobis distance calibrated like the regional score.
This distance is the most efficient of these shower-shape scores and selects $47\%$ and $22\%$ of the two front-face samples.
With signal labels, we train a single gradient-boosted tree classifier~\cite{Pedregosa:2011sklearn} on the $13$ variables of the $40\,000$ photons used neither for calibration nor for testing and of one half of each front-face sample.
It is calibrated like the other scores and evaluated on the other half of each sample.
Over five trainings it selects $(95\pm1)\%$ and $(95\pm2)\%$ of the two samples, compared with $96\%$ and $97\%$ for the regional score.

The classifier on the latent coordinates is a multilayer perceptron with hidden layers of $256$, $128$, and $64$ units, trained separately for each flow and each front-face sample.
Half of the $10^5$ photons of the second sample, drawn from all of its subsets, and half of the signal showers are used for training, and the other halves for evaluation, with the threshold set so that a fraction $10^{-3}$ of the evaluation photons is selected.
Averaged over the five flows and eight random splits, it selects $99\%$ of both front-face samples.

\end{appendix}

\bibliography{SciPost_Example_BiBTeX_File}
\end{document}